 \documentclass[twocolumn,showpacs,preprintnumbers,amsmath,amssymb]{revtex4}

\usepackage{graphicx}
\usepackage{dcolumn}
\usepackage{bm}

\begin{document}

\title{ 
Theoretical analysis of magnetic force microscopy contrast
in multidomain states of magnetic superlattices
with perpendicular anisotropy
}
\author{N. S. Kiselev, I. E. Dragunov, V. Neu, U. K. R\"o\ss ler, A. N. Bogdanov}
\affiliation{IFW Dresden, Postfach 270116, D-01171 Dresden, Germany}
\affiliation{Donetsk Institute for Physics and Technology, 
83114 Donetsk, Ukraine}

\date{\today}

\begin{abstract}
{Recently synthesized magnetic multilayers 
with strong perpendicular anisotropy 
exhibit unique magnetic properties including 
the formation of specific multidomain states.
In particular, antiferromagnetically coupled 
multilayers own rich phase diagrams 
that include various multidomain ground states.
Analytical equations have been derived 
for the stray-field components of these 
multidomain states in perpendicular multilayer systems.
In particular, closed expressions for stray fields 
in the case of ferromagnetic and antiferromagnetic stripes
are presented.
The theoretical approach provides a basis 
for the analysis of magnetic force 
microscopy (MFM) images from this
novel class of nanomagnetic systems.
Peculiarities of the MFM contrast 
have been calculated for realistic tip models.
These characteristic features in 
the MFM signals can be employed for the 
investigations of the different multidomain modes.
The obtained results are applied 
for the analysis of multidomain modes 
that have been reported earlier in the literature 
from experiments on [Co/Cr]Ru superlattices.
}
\end{abstract}

\pacs{
75.70.Cn,
75.50.Ee, 
75.30.Kz,
85.70.Li
}
%

         
\maketitle


%
Magnetic multilayers with strong perpendicular
anisotropy are currently investigated as crucial
elements in magnetic sensors, storage technologies,
and magnetic random access memory 
systems \cite{Maffitt06}.
A large group of them belongs to systems with
antiferromagnetic coupling through non-ferromagnetic interlayers 
(e. g. Co/Ru, Co/Ir, [Co/Pt]Ru, [Co/Pt]NiO superlattices)
\cite{Hamada02,Hellwig03,
Itoh03,Baruth06}.
These nanoscale \textit{synthetic} antiferromagnets 
are characterized by new types of multidomain
states, unusual demagnetization processes
and other specific phenomena 
\cite{Hamada02,Hellwig03,Baruth06}.
In contrast to other bulk and nanomagnetic systems, 
the multidomain states in perpendicular antiferromagnetic multilayers
are determined by a 
strong competition between the antiferromagnetic interlayer 
exchange and magnetostatic couplings \cite{Hellwig03,JMMM04,stripes}.
The remarkable role of stray-field effects
in synthetic antiferromagnets and the peculiarities
of their multidomain states are currently investigated 
by high resolution magnetic
force microscopy (MFM)(for recent examples of 
successful experimental tests on domain theory by MFM see,
e.g. Refs.~\cite{Hellwig03,VN04}).
From the theoretical side, only few results have
been obtained on MFM images 
in antiferromagnetically coupled multilayers, 
mostly by numerical methods
\cite{Hug96,Hamada02,Baruth06}.
Here we present an analytical approach that provides
a comprehensive description of stray-field 
distributions and MFM images in 
multidomain states of these nanostructures.
We show that the stray-field components 
and their spatial derivatives, that are crucial
for an analysis of MFM contrast,
own distinctive features for different multidomain states.
These features allow to recognize
the particular distribution of the magnetization
at the surfaces of domains 
and in the depth of the multilayers. 
The quantitative relations from theory
for the MFM contrast can also serve 
to determine the values of
magnetic interactions, i.e. materials 
parameters of an antiferromagnetic multilayer.
We apply our results for an analysis
of multidomain states observed in  [Co/Pt]Ru
multilayers \cite{Hellwig03}.

\begin{figure}
\includegraphics[width=8.5cm] {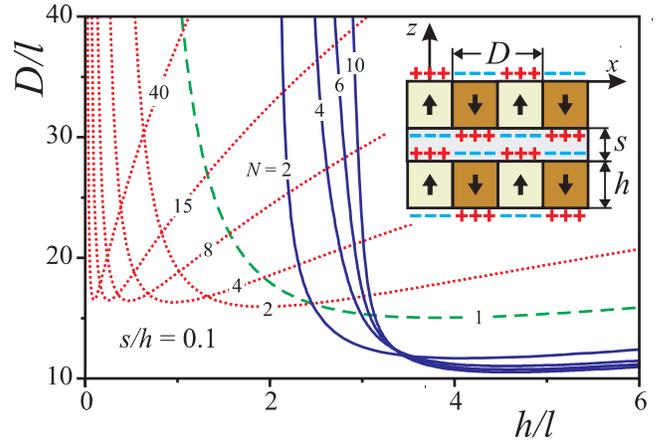}
\caption{
\label{solutions}
(Color online) 
Equilibrium period of stripe states in multilayers.
For various values of multilayer repeat numbers $N$,
the stripe period $D$ 
in units of the characteristic lengths $l$
is shown for the ferromagnetic mode (dotted line),
and antiferromagnetic mode (solid lines)
as function of the magnetic
layer thickness $h/l$ for the ratio of interlayer
to ferromagnetic layer thickness $s/h = 0.1$.
The dashed line indicates the solutions for
a single layer.
The inset introduces the geometrical parameters
of the model.
}
\end{figure}

\begin{figure}
\includegraphics[width=8.5cm] {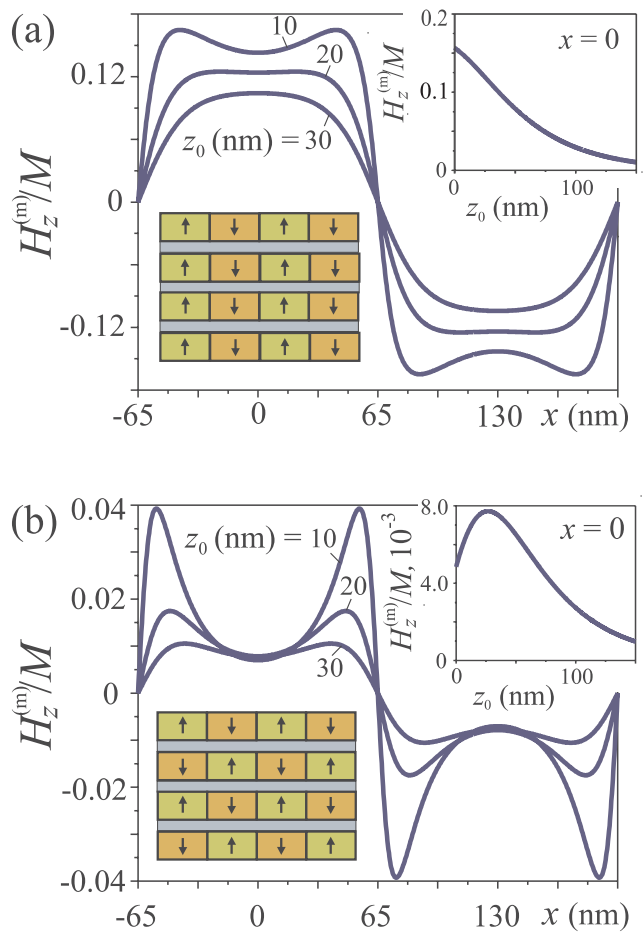}
\caption{
\label{Olafsample}
(Color online)
Calculated stray-field profiles $H_z^{(m)} (x)$
for ferromagnetic (a) and antiferromagnetic (b)
modes for [[Co/Pt]$_7$CoRu]$_4$ multilayers
investigated in \cite{Hellwig03}.
Insets show the perpendicular stray-field 
component $H_z^{(m)} (z_0)$ 
at the center of the stripes
in dependency on the distance $z_0$ above the 
multilayer surface.
In ferro stripes this function
monotonically decreases with increasing $z$, 
while in antiferro stripes it has a maximum 
at a finite distance from the surface.
}
\end{figure}

As a model we consider strong stripes, 
i.e. so-called band domains 
in a superlattice composed of $N$ 
identical layers of thickness $h$ 
separated by spacers of thickness $s$, 
(see, Fig.~\ref{solutions}). 
Note, that the term ``stripe domains'' is also
commonly used to denote multidomain patterns
consisting of stripes with weakly undulating 
magnetization which, however, stays predominantly 
in the layer plane \cite{Hubert98}.
On the other hand, the term band domains is used
to describe structures of homogeneous domains 
with perpendicular magnetization that alternates
between up and down direction. 
These two types of stripe domains should
not be confused.
In the model for an extended multilayer film, 
the multilayer is taken to be infinite in $x-$ and $y-$ directions.
The stripes with alternating magnetization $\mathbf{M}$ 
along the $z$ direction 
and with $|\mathbf{M}|= M \equiv $~const have 
the period length $D$ and are separated by domain 
walls of vanishingly small thickness 
(Fig. \ref{solutions}). 
However, domain walls contribute a positive excess energy
with an area density $\sigma$, which is
included in the model as one of the materials parameters.
The ferromagnetic layers are 
\textit{antiferromagnetically}
coupled via a non-ferromagnetic spacer.
This interaction imposes antiparallel
orientation of the magnetic moments
in adjacent layers, while magnetostatic
forces favor parallel orientation of
the magnetization.
As a results of this competition three
different ground states can be realized
depending on the materials
and geometrical parameters of the multilayer
\cite{stripes}.
Namely, the homogeneous antiferromagnetic state, 
and stripe domains with parallel or antiparallel
arrangement of the magnetization in adjacent
layers (Fig. \ref{Olafsample}). 
We denote these latter two multidomain modes 
as \textit{ferro} and
\textit{antiferro} stripes. 
The energy density for both types of 
stripe phases can be written
as 
\begin{eqnarray}
w_N (D) = 
 \frac{2 \sigma }{D}
\pm\frac{J}{h}\left(1-\frac{1}{N}\right)  +2 \pi M^2\, w_m (D)\,.
\label{energy0}
\end{eqnarray}
The first term in $w_N$ is the domain wall energy,
$J>0$ is the antiferromagnetic exchange interaction 
via the spacer layer. 
The stray field energy $w_m$ 
can be expressed 
as a set of finite integrals \cite{FTT80,stripes},
\begin{eqnarray}
\label{energyStripeMC}
w_m (D)  =  1+ \Omega (D,h) 
\mp \frac{1}{N} \sum_{k=1}^{N-1} (N-k)\,\Xi_k (D,h,s) ,
\end{eqnarray}
where  $T = h+s$ is the period length of the superlattice,
\(\Xi_k  = 2\, \Omega( D,T k)
\,-\,\Omega(D,T k +h) 
\,-\,\Omega(D, T k -h)\), and
\begin{eqnarray}
\label{Omega}
\Omega (D, \omega)= \frac{8 \, \omega^2 }{\pi h D} \int_0^1 (1-t)\ln 
\left|\tanh \left(\frac{\pi \omega t}{D} \right)\right|d t\,.
\end{eqnarray}
The upper (lower) sign in Eqs.~(\ref{energy0}), and in 
(\ref{energyStripeMC}) for $w_m$, corresponds 
to ferro(antiferro) stripes.
The equilibrium domain configuration
of the stripes is derived by minimization 
of $w_N$ with respect to the stripe period $D$ \cite{stripes}.
Introducing a new length scale, based on
the \textit{characteristic length}
$l = \sigma/(4 \pi M^2)$, one can
express the solutions for
reduced periods $D/l$ as functions
of three parameters, 
the reduced geometrical sizes 
($h/l$), ($s/l$),
and the repeat number of the multilayer $N$.
Typical solutions $D/l$ as functions of
$h/l$ are presented in Fig.~\ref{solutions}.
The solutions for ferro stripes exist
for arbitrary values of $h$ and $s$.
In the two limiting cases of large 
and small spacer thickness, 
the solutions asymptotically approach 
the behavior of the known solutions 
for individual layers  (see Ref.~\cite{Malek,Suna85})
with a thickness $h$ for the case  $s \gg h$
and with effective thickness $hN$ for $s \ll h$.
The solutions for antiferro stripes with even
$N$ exist only in an interval bounded from below, 
$ h > h_{cr} (N)$.
The period $D$ tends to infinity 
at a critical thickness $h_{cr}$.
Strictly speaking,
for odd $N$ antiferro stripes 
(similarly to ferro stripes) exist 
for arbitrarily small layer thickness $h$.
However, their periods increases so steeply
that a single domain state is practically
reached when the period exceeds the 
lateral size of the layer. 
The calculated periods for odd-numbered
multilayers has similar have similar size 
as those for antiferro stripes 
with even $N$.
The phase diagrams 
considering these stripe ground-states 
show that both types of stripes can exist 
as stable or metastable states
in extended and overlapping 
ranges of the material parameters \cite{stripes}.
The extended co-existence regions of different
types of multidomain states in the phase diagrams 
also entails the possibility 
to create complex ``interspersed'' patterns 
that consist of subdomains with ferro 
and antiferro stripes.
For identical values of the materials
parameters the equilibrium domain widths for ferro 
and antiferro stripes can differ considerably (see Fig. \ref{solutions}).
Hence, the \textit{mixed} stripe patterns 
can include regions with different domain sizes.
Exactly such structures have been 
observed in some  [Co/Pt]Ru 
multilayers \cite{HellwigPrivate}.
In addition to the differing characteristic periods
of ferro and antiferro stripes, these stripe patterns
also cause different distributions of the stray
fields $\mathbf{H}^{(m)}$ at the sample 
surfaces.
The stray-fields can be probed by 
magnetic force microscopy, however, the properties
of the stray fields peculiar to the different stripe
patterns are rather subtle. 
Thus, the experimental observation 
and quantitative evaluation of these differences 
must be based on a detailed comparison 
with theoretical model calculations.
Convenient analytical expressions for
stray field components and their spatial derivatives 
for multilayers with ferro and antiferro stripes
can be derived 
from Eqs.~(\ref{energy0})--(\ref{energyStripeMC}).
The mathematical equivalence 
between electro- and magnetostatic
problems allows to
treat a magnetized layer as 
a system of ``charges'' distributed
over its surfaces \cite{Hubert98}.
The  stray field above the sample surface 
can be expressed as a superposition 
of the stray fields from the 2$N$ interface 
planes with ``charged'' stripes (Inset in Fig. \ref{solutions}).
By solving the magnetostatic problem for a plane with 
``charged'' stripes (see Appendix) one can 
write the following solutions for the stray 
field components $\mathbf{h}^{(m)} (x, z)$
\begin{eqnarray}
h_x^{(m)}(x,z)  =  2M \ln \left| \frac{\cosh \left(2\pi z/D \right)
 - \sin \left(2 \pi x/D \right)}
{\cosh  \left(2\pi z/D \right) 
+ \sin  \left(2\pi x/D \right)}\right|,
\label{strayfields1x}
\end{eqnarray}
\begin{eqnarray}
h_z^{(m)}(x,z)  =
4M \underbrace{\arctan \left[ \cos  \left(2\pi x/D \right)/
\sinh  \left(2\pi z/D \right) \right]}_{\upsilon_0(x,z)}.
\label{strayfields1z}
\end{eqnarray}
Then the total stray field of the multilayer at 
a distance $z_0$ above the surface
can be written as
\begin{eqnarray}
\mathbf{H}^{(m)} (x, z_0)= \sum_{k=0}^{N-1}
\Gamma_k \left[ \mathbf{h}^{(m)} (x, z_0 +T k)\right.
\nonumber\\
\left.- \mathbf{h}^{(m)} (x, z_0 +Tk+h)\right]\,.
\label{strayfield2}
\end{eqnarray}
The factor $\Gamma_k = (-1)^k$ holds for antiferro stripes,
and $\Gamma_k = 1$ for ferro stripes.
Spatial derivatives of $H_z^{(m)}$ with respect 
to $z$ are important for the analysis of the MFM images.
The derivative $\Upsilon_{n} (x, z_0) =\partial^{n}H_z^{(m)}/{\partial z^{n}} $can be 
derived analytically by differentiation 
of Eq.~(\ref{strayfield2}) as
\begin{eqnarray}
\Upsilon_{n} (x, z_0) = 4M \left(\frac{2 \pi}{D}\right)^{n}\ \ \ \ \ \ \ \ \ \ \ \ \ \ 
\nonumber\\
\times\sum_{k=0}^{N-1} \Gamma_k
\left[ \upsilon_{n} (x, z_0 +T k) 
- \upsilon_{n} (x, z_0 +Tk+h)\right]\,.
\label{strayfield3}
\end{eqnarray}
We introduce here a set of functions $\upsilon_{n} (x,z)$
which are derivatives of the function $\upsilon_{0} (x,z)$
defined in Eq.~(\ref{strayfields1z})
with respect to the normalized geometry parameter $\xi = 2 \pi z/D$,
\begin{eqnarray}
\upsilon_{n} (x,z) \equiv  \partial^n \upsilon_0/\partial \xi^n =
\cos \left( 2 \pi x/D\right) \frac{G_n (x,z)}{g_{+}^{n} (x,z)}, 
\label{strayfields4}
\end{eqnarray}
where
\begin{eqnarray}
& & g_{\pm} = \left[\cosh \left( 4 \pi z/D\right)
 \pm \cos \left( 4 \pi x/D\right) \right]/2,
\nonumber \\
& &  G_1 = - \cosh \left( 2 \pi z/D \right), \quad
\nonumber \\
& & G_2 = \sinh \left(  2 \pi z/D \right)(1+g_{-}), \;
 \nonumber \\
& & G_3 = - \cosh \left(  2 \pi z/D \right)(2 g_{-}^2 -g_{+}^2+2g_{+}-2),
\nonumber \\
& & G_4 = \sinh \left( 2 \pi z/D \right)
\nonumber \\
& & \ \ \ \ \ \ \ \: \times\left[ 6(g_{-}+1)^2(g_{-}-1)+g_{+}^2(1 -5g_{-})\right]\,.
 \nonumber \\
\label{strayfields4b}
\end{eqnarray}
Together with the equation $d w_N /d D =0$, 
which determines the equilibrium domain
period, Eqs. (\ref{strayfield2}) and (\ref{strayfield3}) 
describe the stray field and its spatial
derivatives as a function of the coordinates $x$, $z$ 
for a multilayer in a stripe state.
The stray field $\mathbf{H}^{(m)}(x,z_0)$ (\ref{strayfield2})
and the derivatives  $\Upsilon_{n} (x, z_0)$
(\ref{strayfield3})
are expressed as sets of analytical functions
(\ref{strayfields1x}), (\ref{strayfields1z})
(\ref{strayfields4}), and (\ref{strayfields4b}).
The expressions depend on the geometrical parameters
and, via the equilibrium domain widths $D$, on the 
material parameters of the multilayer.
These analytical expressions can be readily evaluated
by elementary mathematical means.

In order to demonstrate the main features of
the stray fields $H_z^{(m)} (x,z)$ 
from the ferro and antiferro stripes
we evaluate these functions for
a multilayer 
[[Co/Pt]$_7$CoRu)]$_4$
with magnetic and geometrical
parameters corresponding
to a sample that was
investigated experimentally in \cite{Hellwig03}.
In this superlattice 
the ferromagnetic constituents 
are magnetic [Co/Pt]$_7$ multilayers 
with thicknesses of the ferromagnetic Co-layers 0.4~nm 
and thickness of the Pt layer 0.7~nm. 
The non-ferromagnetic Ru spacer
has a thickness of 0.9 nm and mediates an indirect 
antiferromagnetic interlayer exchange.
The domain period
has been determined as $D$ = 260~nm
\cite{Hellwig03}.
For ferro and antiferro stripe modes
the functions $H_z^{(m)} (x,z)$ are markedly
different both in the intensity and
in the location of characteristic extremal points
(Fig.~\ref{Olafsample} ).
Moreover they display
\textit{qualitatively} different
functional dependencies on the
distance from the multilayer surface $z_0$ 
(see Insets in Fig.\ref{Olafsample}).

The stray-field distribution over the
multilayer surface can be viewed as
a superposition of magnetostatic fields from 
systems of ``charged'' bands.
This allows to give a simple physical
interpretation for the main 
features of the stray field  
profiles in Fig. ~\ref{Olafsample}.
First we consider ferro stripes.
It is convenient to separate
the total stray-field over 
a domain to two contributions:
one created by the top and bottom
bands of the domain("self" field), 
and that produced by all other "charged"
bands.
In the domain centers 
near the sample surface,
$z_0 \ll D$,
the stray ``self'' field is
small due to screening effects
of the domain top and bottom surfaces.
Because the bands change their ``charges''
at the domain walls
the stray field is
substantially enhanced
above the walls.
As a result, 
the profiles $H_z^{(m)} (x,z)$ 
have  characteristic wells
in the domain centers
for $z_0 \ll D$.
For increasing distance $z_0$ from the surface 
the difference between
values of $H_z^{(m)}$
in the center and at the domain
edges decreases
due to the increasing influence of 
neighbouring poles.
For large distances $z_0$ the wells
disappear and the profiles obtain 
a typical bell-like shape 
(compare the traces for $z_0$ = 10 and 30~nm 
in Fig.~\ref{Olafsample} (a)).
The antiferro stripe mode can
be obtained from those for ferro stripes
by changing the  magnetostatic ``charges'' 
for the bands in all even layers.
This weakens the total stray field
and sharpens the difference between
the stray fields at the center
and near the domain edges.
Finally, the competing character of 
the stray field contributions 
from odd and even layers causes 
the nonmonotonic dependence of $H_z^{(m)}(z)$
(Inset in Fig. ~\ref{Olafsample} (b)).

\begin{figure}
\includegraphics[width=9cm] {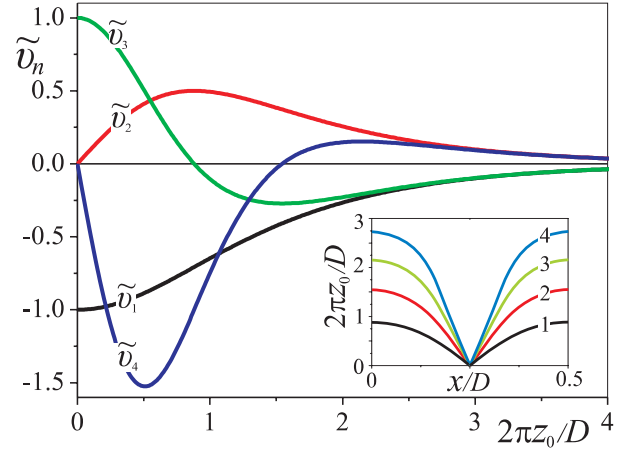}
\caption{
\label{upsilons}
(Color online)
Characteristic functions 
$\tilde{ \upsilon}_n (z_0) \equiv \upsilon_n(0,z_0)$
(Eq. (\ref{strayfields7}))
describe MFM images
in the center of domains.
Inset indicates the location
of the extrema of the functions
$ \upsilon_n (x,z_0)$ in the $xOz$ plane.
}
\end{figure}

Generally the functions $\mathbf{H}^{(m)}(x,z_0)$,
$\Upsilon_{n}^{(F)}(x,z_0)$,
$\Upsilon_{n}^{(AF)}(x,z_0)$
have a number of characteristic features 
that can be utilized in new methods
to investigate the multidomain modes.
One method can be based on measuring
the MFM contrast in the center of the domains.
In this case
$x = k D,\, \, k=0,1,2,3...$ and the functions 
$\tilde{\upsilon}(z) \equiv\upsilon_{n} (kD,z)$ 
are reduced to the following expressions
\begin{eqnarray}
& & \tilde{\upsilon}_{0} (z)= \arccos \left[\tanh \left(
2 \pi z /D \right) \right]
, \quad
\nonumber \\
& & \tilde{\upsilon}_{1} (z)= - \cosh^{-1} 
\left( 2 \pi z /D \right),
\nonumber \\
& & \tilde{\upsilon}_{2} (z)=  \sinh \left( 2 \pi z/D
 \right)\cosh^{-2} \left( 2 \pi z /D \right), 
 \quad
\nonumber \\
& & \tilde{\upsilon}_{3} (z)= - (1-\sinh^2 \left( 2 \pi z
/D \right))\cosh^{-3} \left( 2 \pi z /D \right), 
\nonumber \\
& & \tilde{\upsilon}_{4} (z)=  \sinh \left( 2 \pi z/D \right) 
\frac{\sinh^2 \left( 2 \pi z/D \right) -5}{\cosh^{4} \left( 2 \pi z /D \right)}.
\label{strayfields7}
\end{eqnarray}
Profiles $\tilde{\upsilon}_{n} (z_0)$ from Eqs.~(\ref{strayfields7})
for $n$ = 1,2,3,4 are plotted in Fig.~\ref{upsilons}.
Their characteristic features
can be used to ascertain the type of the stripe mode 
and should even allow quantitative evaluation of 
magnetic properties in multilayers, 
if and when they are accessible in experiment.

Owing to the properties of the magnetic probe, 
a signal from a magnetic force microscope 
generally differs strongly 
from the $H_z(x)$ profile.
In order to compare the expected outcome 
of different domain configurations, 
the MFM contrast has to be calculated for 
realistic tip models. 
The MFM signal for a magnetic cantilever oscillating 
in \textit{z}-direction is given by \cite{Grutter92,Lohau99}
\begin{eqnarray}
\label{phase}
\Delta \Phi & =&  -\frac{Q}{k} \left(\frac{\partial F_z}{\partial z}\right) \nonumber \\
            & =&  -\frac{Q}{k} 
\mu_0  \int_{tip} \frac{\partial^2 \left[\mathbf{M}^{(tip)}(\mathbf{r}) 
\cdot \mathbf{H}^{(m)} (\mathbf{r}) \right]}{ \partial z^2}\;dV
\,.
\end{eqnarray}
Here, $\Delta \Phi$ is the measured phase shift between 
excitation and oscillation due to 
the force gradient $\partial F_z/\partial z$
that acts on the cantilever in the stray field of the sample 
$\mathbf{H}^{(m)} (\mathbf{r})$. 
$Q$ and $k$ are the quality factor of the oscillation and 
the spring constant, respectively. 
Assuming a rigid tip magnetized in \textit{z}-direction, 
i.e., a tip with a homogeneous magnetization distribution,
$\mathbf{M}^{(tip)} \equiv$~const
throughout the tip volume, that does not change during 
the scan across the stray field of the sample, 
the expression for the force gradient simplifies to
\begin{eqnarray}
\label{phase2}
\frac{\partial F_z}{\partial z}=
M_{z}^{(tip)} 
\int_{tip} 
\frac{\partial^2  H_{z}^{(m)}(\mathbf{r})}{ \partial z^2}\; dV\,.
\end{eqnarray}
The volume integration is a crucial step as it modifies 
the signal compared to the profile 
estimated by the second stray field derivative.

\begin{figure}
\includegraphics[width=8.5cm] {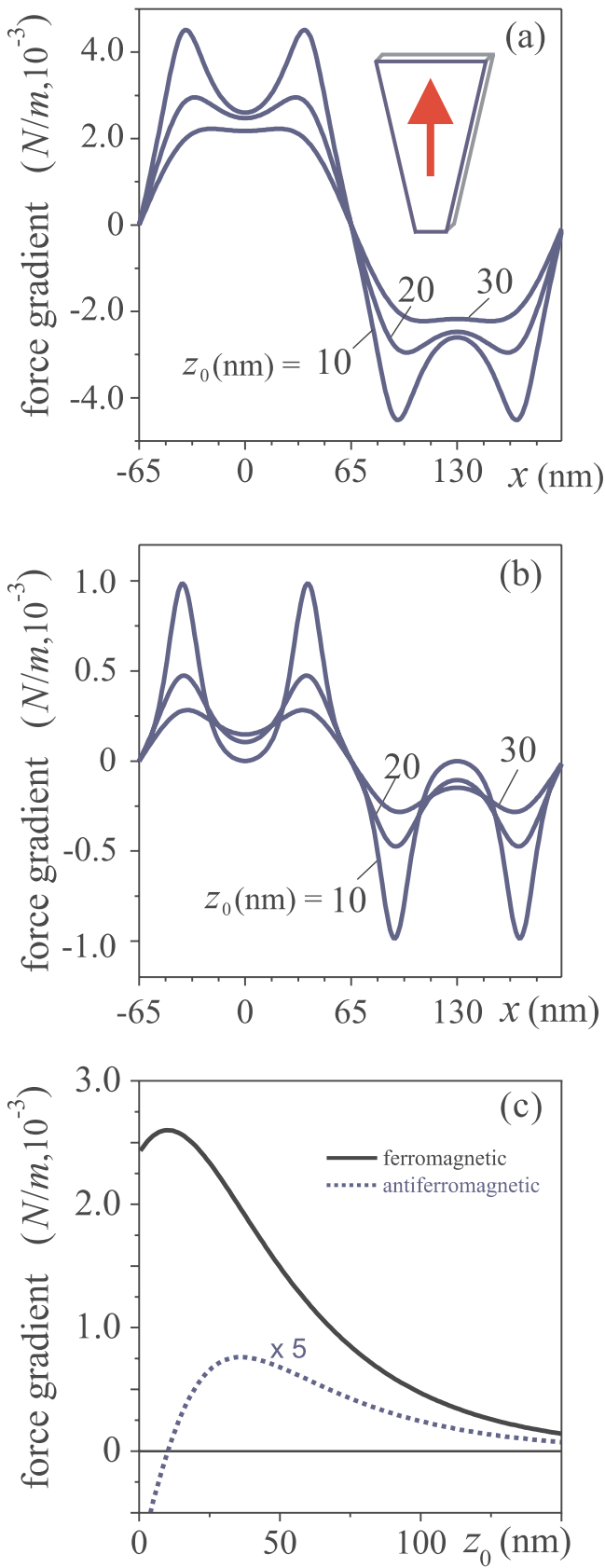}
\caption{
\label{MFMcontrast}
(Color online)
Calculated force gradient profiles for
a 4-sided pyramidal tip and ferro (a),
antiferro (b) modes in the center of domains as
function of the distance $z_0$ from the surface (c)
for  [[Co/Pt]$_7$CoRu]$_4$ 
multilayers with the same geometrical parameters
as in Fig.~2.
}\end{figure}

A realistic tip geometry can be modelled by a truncated 
triangle placed in the \textit{x-z}-plane
(see inset in Fig. \ref{MFMcontrast}). 
This mimics the two parallel sides 
of the 4-sided pyramidal geometry of a typical MFM tip. 
The two-dimensional model simplifies considerably 
the calculations. The error incurred by 
this reduced model is minor because 
of the infinite extension of the domain models
in the \textit{y}-direction. 
\begin{figure}
\includegraphics[width=8.0cm] {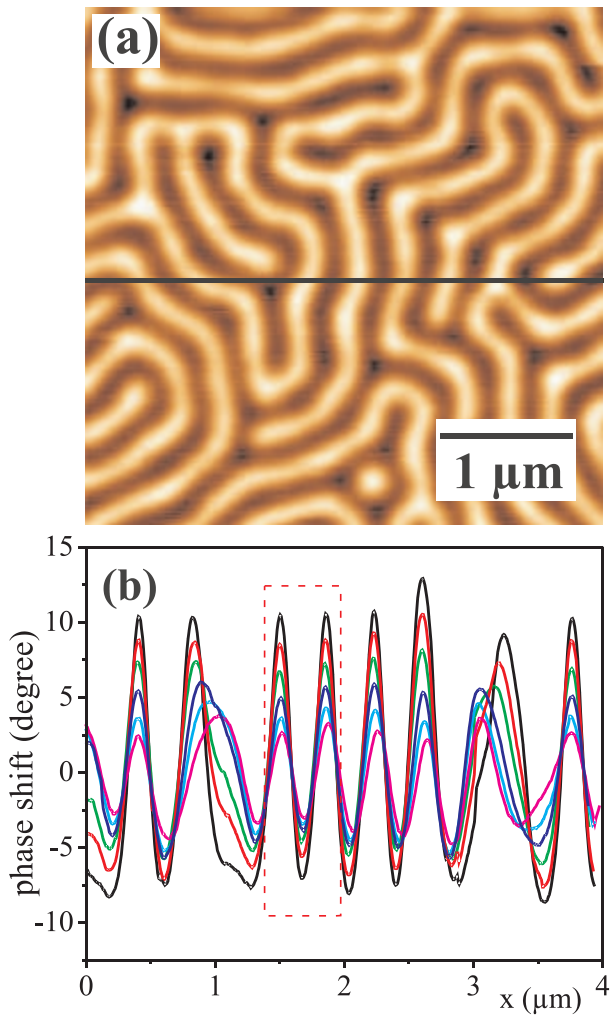}
\caption{
\label{stripe18}
(Color online)
Maze domain pattern in [(Co/Pt)$_8$)CoRu]$_{18}$
observed by MFM at room temperature (a) 
and line scans along the
marked line for  $z_0$ 
= 10, 20, 30, 40 ,50 and 60 nm.
}\end{figure}

To demonstrate this approach we analyze the MFM 
contrast measured across a thick [(Co/Pt)$_8$)CoRu]$_{18}$ 
multilayer prepared at Hitachi GST (for details on these
multilayers, see Refs.~\cite{Hellwig03}).
The MFM pictures show a typical 
maze pattern of ferro stripes 
with perpendicular magnetization 
at room temperature Fig.~\ref{stripe18}~(a)). 
For comparison with contrast calculations
the MFM signal along the marked line of 
the image was recorded repeatedly 
(with the slow scan axis disabled) 
at lift heights of 10, 20, 30, 40, 50 and 60 nm. 
Averaged scan lines for each lift height 
in Fig.~\ref{stripe18}~(b) 
display a reduced contrast for increasing scan height. 
\begin{figure}
\includegraphics[width=8.5cm] {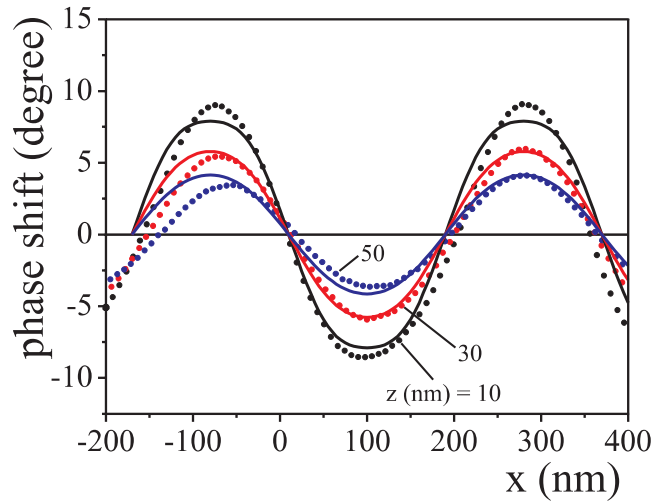}
\caption{
\label{MFM18}
(Color online)
Experimental (points) and
calculated (solid lines)
force gradient profiles
[(Co/Pt)$_8$)CoRu]$_{18}$
for MFM scans
corresponding to the framed
area in Fig.~\ref{stripe18}(b).
}\end{figure}

The very regular domain pattern in the center 
of this scan (framed area) was modeled according 
to Eq.~(\ref{strayfield3}) as parallel FM stripes 
with the period of 360 nm 
based on an $N=18$ multilayer 
with the known layer architecture.
With the resulting second stray field derivatives 
the MFM phase shift was computed according to 
Eq.~(\ref{strayfield3}) using the above mentioned tip model. 
As cantilever parameters a spring constant of $k$ = 2 N/m 
and a quality factor of $Q$ = 100 were used. 
The height, pyramid angle and tip apex were chosen 
as 4~$\mu$m, 30 and 50 nm, respectively, and 
the film coating was assumed to be 30~nm. 
As an adjustable parameter the tip magnetization 
$M_z$,tip was set to 3$\cdot$10$^{6}$  A/m. 
With this reasonable value the computed phase 
shift magnitude, line profile and lift height 
dependency compare very well with the experimental 
data (Fig. \ref{MFM18}).
Such calculations can thus be used to predict 
differences in the MFM contrast of 
the two distinct types of stripe domains. 

Fig.~\ref{MFMcontrast} shows the calculated force gradient 
profiles for the ferro and antiferro stripes presented 
in  Fig.~\ref{Olafsample}. 
The profiles reflect general features 
of the stray-field distributions in the multidomain patterns.
Here as well, quantitative and qualitative differences 
allow to distinguish ferro and antiferro stripes.
The signal from the antiferro stripes is clearly weakened,  
but it is large enough to be measured in a typical MFM setup 
which allows the detection of a few 10 $\mu$N/m (10$^{-2}$ dyn/cm)
\cite{Schendel00}.
However, as quantitative MFM measurements are still 
rare and require precise calibration routines
\cite{Schendel00,Lohau99} the absolute value of the signal 
is not a reliable criterion for the distinction between 
different stripe configurations. 
More importantly, 
the force gradient directly 
above the center of a domain
shows a monotonically decreasing 
signal strength for increasing 
scan height $z_0$ in case of the ferro stripes. 
Above the antiferro stripes, on the other hand, 
the force gradient is increasing with increasing 
scan height, at least in the range from 10 to 30 nm.
This qualitative difference is demonstrated again 
in Fig.~\ref{MFMcontrast}, where 
the force gradient experienced 
by a realistic tip model is plotted 
as a function of $z_0$ for the two cases. 
The non-monotonic behavior of the force gradients 
observed above a multidomain structure 
can be taken as a clear fingerprint
of an antiferro stripe state
within a multilayer stack.
The calculation even reveals a sign change as 
an additional signature of the antiferro stripes, 
but this appears at a scan height smaller 
than 10~nm, which is experimentally very difficult 
to access, as topographic information can influence 
the measurement. 

The solutions for the stray field (\ref{strayfield2})
and its derivatives (\ref{strayfield3})
can be simplified in the practically important case of
large domains, $D \gg h$.
Expansion of $\Upsilon_n (x,z_0)$ 
with respect to the small parameter
$h/D  \ll 1$ yields
for ferro stripes
\begin{eqnarray}
\Upsilon_{n}^{(F)} (x,z) = -4M N \left(2 \pi/D\right)^n 
\;\upsilon_{n+1} (x,z), 
 \label{strayfields5f}
\end{eqnarray}
and for antiferro stripes with even $N$
\begin{eqnarray}
\Upsilon_{n}^{(AF)} (x,z) = -2M N\!
\left(2 \pi/D \right)^n 
(1\!+\!s/h) \upsilon_{n+2}(x,z). 
 \label{strayfields5af}
\end{eqnarray}
For antiferro stripes in multilayers
with odd $N$ the functions $\Upsilon_{n} (x,z)$
are given by Eq.~(\ref{strayfields5f}).
Note that the functions
$\Upsilon_{n}^{(F)}$ from the Eq.~(\ref{strayfields5f}) 
are proportional to $\upsilon_{n+1}$,
while the functions $\Upsilon_{n}^{(AF)}$
in Eq.~(\ref{strayfields5af})
can be expressed by derivative of the functions $\upsilon_{n+1}$
through 
the relation $\upsilon_{n+2}= \partial \upsilon_{n+1} / \partial z$.
This means that in this limit of large domains,
the functions $\Upsilon_{n}$
in the antiferro mode behave as $z$-derivatives
of the corresponding functions of ferro modes.
In particular for $n=0$, the Eqs.~(\ref{strayfields5f})
and (\ref{strayfields5af})
give the perpendicular stray field components
for ferro stripes and antiferro stripes with
odd $N$,  correspondingly, by the expressions
\begin{eqnarray}
& & H_z^{(F)} (x,z_0)= - 4M N \;\upsilon_1(x,z_0), 
\nonumber\\
& & H_z^{(AF)} (x,z_0)= - 2M N (1+s/h)\;\upsilon_2(x,z_0).
 \label{strayfieldFAF}
\end{eqnarray}
%
%
%

In conclusion, we have presented analytical solutions for
the stray field (Eq.~(\ref{strayfield2})) 
and its spatial derivatives (Eq.~(\ref{strayfield3}))
in multidomain states of
 magnetic multilayers with out-of-plane magnetization.
These solutions can be applied to calculate
the MFM contrast for a realistic tip geometry
using Eq.~(\ref{phase2}).
It is shown that the ground-state ferro and antiferro stripe
structures in antiferromagnetically
coupled multilayers differ by their period lengths 
and by the spatial distribution of their stray-fields. 
Our analytical calculations executed within a simplified
model of one-dimensional multidomain patterns with fixed
magnetization orientation and infinitely thin domain
walls are able to reproduce the general features of 
MFM images from antiferromagnetically coupled multilayers.
These features can be used to identify
different types of multidomain patterns and extract
values of the magnetic interactions.
They also create a basis for more 
detailed investigations on more realistic models.
Such models should consider 
distortions of the magnetization with tilting away from 
the perpendicular direction in sizeable
fractions of the domains 
due to the finite strength of uniaxial anisotropy,
or a finite width of the domain walls 
between domains combined with the appearance of 
magnetic charge distributions at the walls.
Additionally, it was recently shown 
that ferro stripes are unstable with respect 
to a lateral shift of domains
in adjacent layers\cite{APLshift}.
This creates specific multidomain patterns 
with distinct modulations across the multilayer stack.
Such effects can be consistently taken 
into account by numerical solutions of 
the micromagnetics equations for specific cases.
Here, we have analyzed the backbone structure for all
of these types of spatially inhomogeneous 
distributions of the magnetization within the multilayer 
and stray fields over its surface.



\appendix
\begin{center}
\textbf{Acknowledgments}
\end{center}


The authors thank
O. Hellwig, J. McCord, A.~T.~Onisan, 
and R. Sch{\"a}fer for helpful discussions.
We also thank O. Hellwig (Hitachi GST)
for providing the Co/Pt/Ru multilayer sample
and C. Bran for performing the MFM measurements.
Work supported by DFG through SPP1239, project A8.
N.S.K., I.E.D, and A.N.B.\ thanks H.~Eschrig for support 
and hospitality at IFW Dresden. 


\appendix
\begin{center}
\textbf{Appendix}
\end{center}


%
The scalar potential of a sheet with ``charged''
stripes (Inset Fig.~\ref{solutions})
can be derived by solving Poisson's equation
(see, e.g., \cite{Malek,Suna85})
\begin{eqnarray}
\phi (x, z) = \frac{4MD}{\pi} 
\ \ \ \ \ \ \ \ \ \ \ \ \ \ \ \ \ \ 
\nonumber\\
\times\sum_{n=1}^{\infty} \frac{\sin(\pi n/2)}{n^2} 
\cos \left( 2\pi n x/D \right) \exp \left(- 2 \pi n z/D \right)\,.
\label{potential2}
\end{eqnarray}
Using the identity 
$n^{-m} = [(m-1)!]^{-1} \int_0^{\infty} t^{m-1} e^{-nt}dt$ 
the Eq.~(\ref{potential2}) can be transformed 
into the following form \cite{FTT80,stripes}
\begin{eqnarray}
\phi (x, z) = \frac{8M\!D}{\pi} \!\int\limits_0^{\infty}\!
 \frac{ \cos (2 \pi x/D) \cosh(t+2 \pi z/D)t dt}
 {\cosh^2(t +2 \pi z/D) - \sin^2 (2 \pi x/D)}\;.
 \nonumber\\
\label{potential3}
\end{eqnarray}
From this closed expression, the components of the
stray field $\mathbf{h}^{(m)} = - \nabla \phi$
are readily derived in the analytical form given by
Eqs.~(\ref{strayfields1x}) and (\ref{strayfields1z}).




\end{document}